\documentclass{jfm}

\usepackage{tikz,graphicx,color,amsmath,amssymb}
\usepackage{natbib}
\usepackage{xcolor}
\definecolor{mygreen}{rgb}{0 .5 0}

\newcommand{\RomanNumeralCaps}[1]
\linenumbers

\usepackage{tabularx}
\usepackage{overpic}
\usepackage{indentfirst}
\usepackage{relsize}
 
\usepackage[colorlinks]{hyperref}
\hypersetup{
	linkcolor = blue,
	urlcolor   = blue,
	citecolor  = black,
}

\usepackage[nameinlink]{cleveref}
\crefname{section}{section}{sections}
\crefname{subsection}{subsection}{subsections}
\crefname{figure}{figure}{figures}
\crefname{table}{table}{tables}
\crefname{equation}{}{}
\Crefname{section}{Section}{Sections}
\Crefname{subsection}{Subsection}{Subsections}
\Crefname{figure}{Figure}{Figures}
\Crefname{table}{Table}{Tables}

\newcommand{\bu}{\mathbf{u}}

\newcommand{\beq}{\begin{equation}}
\newcommand{\eeq}{\end{equation}}

\title{Domain-filling rolls in two-dimensional fixed-flux Rayleigh--B\'enard convection}

\author{Mathew D.\ B.\ Lewis, Zehan Zheng \and David Goluskin
\corresp{\email{goluskin@uvic.ca}}}

\affiliation{Department of Mathematics and Statistics, University of Victoria, Victoria, BC, V8P 5C2, Canada}

\begin{document}
\maketitle

\begin{abstract}
\noindent 
Rayleigh--B\'enard convection of large horizontal extent sometimes self-organizes into domain-filling structures. In two dimensions, domain-filling rolls persist---from some but not all initial conditions---when velocity boundary conditions are stress-free. When velocity boundary conditions are no-slip, domain-filling rolls are not found with fixed-temperature thermal boundary conditions, but they have not been sought with fixed-flux thermal boundary conditions. Here we explore the latter missing case, which is hard to predict because no-slip boundaries do not encourage domain-filling rolls, but fixed-flux boundaries give domain-filling structures in three dimensions for either boundary condition on velocity. We simulate convection with fixed-flux, no-slip boundaries in two-dimensional domains with horizontal period 20 times their height and at various combinations of the fixed-flux Rayleigh number $R$ and Prandtl number $Pr$. Domain-filling rolls, which are easily found when they are weakly nonlinear at small $R$, are continued to other $(R,Pr)$ by changing these parameters slowly in time. The $(R,Pr)$ regime where we find domain-filling rolls is substantial but has a boundary. When $R$ is too large or $Pr$ too small, relative to each other, a domain-filling roll pair breaks up into two pairs. These findings contrast with other combinations of dimension and boundary conditions, where scale selection has not been seen to depend strongly on parameter values. The present case helps disentangle competing effects of dimension and boundary conditions, and it offers a more stringent test for explanations of scale selection that have been proposed.
\end{abstract}


\section{Introduction}
\label{sec:intro}

Rayleigh--B\'enard convection (RBC) in a fluid layer can be sustained by various thermal boundary conditions, provided that heat flows upward across the layer on average. Most common is the fixed-temperature case, where the bottom boundary is fixed to a higher temperature than the top one. Also common is the fixed-flux case, where there is a uniform heat flux inward at the bottom and outward at the top. With these boundary conditions or others, as thermal forcing is increased via the dimensionless Rayleigh number $R$, flow structures such as thermal plumes typically become smaller and more numerous. At the same time, these small-scale structures self-organize into superstructures whose characteristic scale is comparable to the layer height or larger. The nature of these superstructures can vary greatly between Rayleigh--B\'enard configurations with different boundary conditions on the temperature or the velocity, and between two-dimensional (\mbox{2-D}) and three-dimensional (\mbox{3-D}) flows.

In \mbox{3-D} RBC of large horizontal extent, the superstructures that emerge depend starkly on the thermal boundary conditions. This is shown by the direct numerical simulation (DNS) study of \citet{Vieweg_2021a} in a horizontally periodic domain with both periods 60 times the layer height. They imposed four different sets of boundary conditions, always the same at the top and bottom: thermal conditions were either fixed-temperature or fixed-flux, and velocity conditions were either no-slip or stress-free. In the fixed-temperature cases with either velocity condition, the pattern of superstructures evolved slowly in time but maintained a dominant horizontal scale of several times the layer height, as found also in earlier DNS with no-slip boundaries \citep{VonHardenberg2008,Pandey_2018,Krug_2020}. However, in the fixed-flux cases with either velocity condition, \citeauthor{Vieweg_2021a} found that superstructures initially looked similar to their fixed-temperature counterparts, but they slowly merged until reaching a domain-filling size with all hot plumes coalescing in one upwelling region, and all cold plumes in one downwelling region. This finding does not depend strongly on $R$ or on the dimensionless Prandtl number $Pr$, which is the ratio of viscosity to thermal diffusivity. The key results of \citet{Vieweg_2021a} were confirmed over several decades of $R$ with $Pr=1$, and then by \citet{Vieweg_2024} over four decades of $Pr$ at relatively small $R$. For imperfectly conducting boundaries that essentially interpolate between fixed-temperature and fixed-flux conditions, \citet{Kaufer_2023} reported superstructures of intermediate size. In all of these cases the eventual scale of superstructures---after initial transients that can be very long---seems not to depend on initial conditions. This suggests, but does not prove, that all of these \mbox{3-D} models have a single turbulent attractor at large $R$.

In \mbox{2-D} RBC, superstructures differ from the \mbox{3-D} case both in their nature and in how they depend on boundary conditions. Often \mbox{2-D} superstructures are nearly independent of time once initial transients have passed. In a wide domain there eventually will be a fixed number of convection rolls of approximately equal size, and only their small-scale features will fluctuate. Different initial conditions lead to different numbers of convection rolls, indicating that there are multiple locally attracting states with their own basins of attraction. Each such state can be identified by the mean width-to-height ratio $\Gamma_r$ of a single roll. With fixed-temperature boundaries in both the no-slip and stress-free cases, \citet{Wang_2020b,Wang_2020} searched for \mbox{2-D} states with different $\Gamma_r$ values. They did this by starting DNS from multiple initial conditions resembling convection rolls with various $\Gamma_r$. With no-slip boundaries, stable superstructures were found with only $O(1)$ values of $\Gamma_r$. The particular $\Gamma_r$ values found depend on $R$ and $Pr$ but are all between $2/3$ and $4/3$. If rolls are initially too wide or too narrow for $\Gamma_r$ to lie in the stable range, rolls will split or merge until a stable superstructure is reached. With stress-free boundaries, stable superstructures were found with $\Gamma_r$ ranging from $8/5$ up to the largest value admitted by the horizontally periodic domain---that is, a pair of rolls with $\Gamma_r$ being half the domain's period. The largest domains simulated by \citet{Wang_2020b} varied with $R$ and $Pr$, so their domain-filling rolls had $\Gamma_r$ values between 8 and 64. In the case of stress-free and fixed-flux---rather than fixed-temperature---boundaries, we again expect that rolls can vary from $O(1)$ to very wide. This is based on our own preliminary simulations (not reported here) and the early computations of \citet{Hewitt1980}, but we emphasize that possible scales in this case have not been studied systematically as done in the \mbox{2-D} fixed-temperature cases.

\begin{table}
\begin{center}
\def~{\hphantom{0}}
\begin{tabular}{lcccc}
& \multicolumn{2}{c}{fixed-temperature} & \multicolumn{2}{c}{fixed-flux} \\[-6pt]
& \multicolumn{2}{c}{\underline{\hspace{130pt}}} & \multicolumn{2}{c}{\underline{\hspace{130pt}}} \\
& no-slip & stress-free & no-slip & stress-free \\[5pt]
\mbox{3-D} (single scale)	& $O(1)$ & $O(1)$ & large & large \\
\mbox{2-D} (range of scales)	& $O(1)$ & $O(1)$ to large & (present study) & O(1) to large
\\
\end{tabular}
\caption{Dominant horizontal scales in units of layer height, for superstructures observed in DNS of RBC with moderate-to-large $R$ in \mbox{3-D} \citep{Vieweg_2021a,Vieweg_2024} and in \mbox{2-D} \citep{Wang_2020b,Wang_2020}. Domains are horizontally periodic. Top and bottom boundary conditions are the same, with thermal conditions either fixed-temperature or fixed-flux, and velocity conditions either no-slip or stress-free. `Large' means that superstructures span domains of the largest horizontal extent among existing studies.} 
\label{tab: intro}
\end{center}
\end{table}

\Cref{tab: intro} summarizes the previous findings we have described on superstructures in \mbox{3-D} and \mbox{2-D} convection, in units where the layer height is 1. There are eight cases in total if we require the same boundary conditions at the top and bottom but allow for a choice of two velocity conditions, two temperature conditions, and either \mbox{2-D} or \mbox{3-D} geometry. The \mbox{3-D} situation is simpler since superstructures seem to select a unique scale that is primarily determined by the thermal boundary conditions. In \mbox{2-D} flows, the effect of thermal boundary conditions on scale selection has not been studied by DNS, but the effect of velocity boundary conditions is also significant. In the \mbox{2-D} fixed-temperature cases, no-slip boundaries seem to allow only $O(1)$ rolls, whereas with stress-free boundaries there is no known upper bound on how wide rolls can be. In fact, there is reason to believe that arbitrarily wide rolls can persist in the \mbox{2-D} stress-free case: some initial conditions lead to a horizontally periodic `windy convection' state \citep{Goluskin2014,Wang_2020b,Wang_2023}, which is dominated by mean flow in opposite directions along the top and bottom boundaries, and which thus resembles a wide roll with the up-flow and down-flow regions removed. Windy convection might be the limit of increasingly wide rolls in the \mbox{2-D} stress-free case, but we do not pursue this idea here.

The present study is aimed at the missing case in \cref{tab: intro}: \mbox{2-D} RBC with fixed-flux, no-slip boundaries. It is hard to anticipate whether domain-filling rolls will be found robustly in this case, even with knowledge of the other seven cases. Past \mbox{3-D} results might suggest that domain-filling rolls can be found in \mbox{2-D} with no-slip fixed-flux boundaries. On the other hand, past \mbox{2-D} results in the fixed-temperature cases suggest that the velocity boundary conditions are significant in \mbox{2-D}, and that no-slip conditions discourage wide rolls. In order to disentangle the effects of dimension and boundary conditions on the scales of superstructures, we must determine what scales are possible in the missing case.

Here we report on \mbox{2-D} DNS of the no-slip fixed-flux case in a wide domain, aiming to determine the region of the $R$--$Pr$ plane where domain-filling rolls can persist. We do not attempt to determine what smaller roll sizes can also persist as other local attractors, as has been done in the \mbox{2-D} fixed-temperature cases \citep{Wang_2020b,Wang_2020}, but we take greater pains to seek domain-filling rolls. 

\begin{figure}
\centering 
\includegraphics[width=0.99\textwidth]{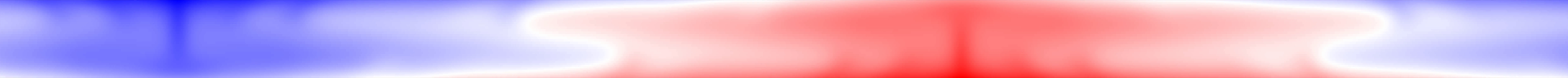}
\caption{Instantaneous temperature field from DNS of a time-dependent but persistent pair of domain-filling rolls in RBC. The domain is horizontally periodic with an aspect ratio of 20. Top and bottom boundaries are no-slip and fixed-flux, and $(R,Pr)=(10^{5},1)$. Fluid is red if hotter and blue if colder. A movie of the evolving flow is available online.}
\label{fig: rolls}
\end{figure}

\Cref{fig: rolls} shows an instantaneous temperature field that is representative of the two-roll states we seek. 
Hotter (red) fluid rises and then flows left or right in the upper part of the domain as it cools; cooler (blue) fluid falls and then flows left or right in the lower part of the domain as it warms. A movie of the time-evolving temperature field is provided.

The rest of this paper is organized as follows. \Cref{sec: comp} describes how we carry out DNS in search of domain-filling rolls. \Cref{sec: res} summarizes our finding, namely the region in the $R$--$Pr$ plane where domain-filling rolls persist. Discussion and conclusions are offered in \cref{sec: con}.

\section{Model and methods}
\label{sec: comp}

The governing equations of RBC are the Boussinesq equations, which model the flow as being divergence-free and having constant kinematic viscosity $\nu$, thermal diffusivity $\kappa$ and coefficient of thermal expansion $\alpha$. We non-dimensionalize by choosing dimensional scales for each coordinate and each field. The length scale is the layer height $h$, so the dimensionless spatial domain is $(x,z)\in[0,\Gamma]\times[0,1]$, where $x$ is the periodic horizontal coordinate and $z$ is the vertical coordinate. The temperature scale $\Delta$ is defined so that $-\Delta/h$ is the dimensional upward temperature gradient at the boundaries, which is set in our case by the fixed-flux boundary conditions. For velocity we use the freefall scale $U_f=\sqrt{g\alpha h\Delta}$, where $g$ is the acceleration of gravity in the downward $-\mathbf{\hat{z}}$ direction, and the time scale is $h/U_f$. With our chosen scales, the Boussinesq equations governing the dimensionless velocity vector field $\bu$, temperature field $T$ and pressure field $p$ are
\begin{subequations}
\label{eq: bouss}
\begin{align}
{\partial_t \mathbf{u}} + \mathbf{u}\cdot\nabla{\mathbf{u}} &= -\nabla{p} + {(Pr/R)^{1/2}} \,\nabla^2\mathbf{u} + T\mathbf{\hat{z}}, \label{eq: nse}\\
 \nabla\cdot{\mathbf{u}} &= 0,\label{eq: incomp}\\
{\partial_t T} + \mathbf{u}\cdot\nabla T &= {(Pr R)^{-1/2}}\,\nabla^2T, \label{eq: T pde}
\end{align}
\end{subequations}
where the fixed-flux boundary condition on $T$ and no-slip boundary condition on $\bu$ are
\begin{align}
\label{eq: BCs}
\partial_zT\big|_{z=0,1}&=-1, &
\bu\big|_{z=0,1}&=\mathbf{0},
\end{align}
and the dimensionless parameters are
\begin{align}
R &= \frac{g\alpha h^3\Delta}{\kappa\nu}, &
Pr=\frac{\nu}{\kappa}.
\label{eq: R Pr}
\end{align}

The parameter $Pr$ in~\cref{eq: R Pr} is the usual Prandtl number, which conveys how quickly the fluid diffuses velocity gradients relative to temperature gradients. The control parameter $R$ is one of two quantities that is called a Rayleigh number in the fixed-flux case. The other fixed-flux Rayleigh number, which we do not use, is an emergent quantity rather than a control parameter. It is obtained by replacing $\Delta$ with the time-averaged dimensional temperature difference between the boundaries \citep{Goluskin_2016c}. This emergent Rayleigh number is valuable for quantitatively comparing fixed-flux and fixed-temperature RBC \citep{Johnston2009}, but here we need only $R$.

Linear stability analysis of the conductive state gives the $R$ values at which different steady roll states bifurcate supercritically. In particular, we solve the linear stability eigenproblem as in section 2.1.1 of \citet{Goluskin_2016c} for fixed-flux no-slip boundary conditions and with horizontal wavenumbers that are integer multiples of $2\pi/\Gamma=\pi/10$. Convection becomes possible as soon as $R$ exceeds $R_L\approx722.6$, where the conductive state becomes linearly unstable, and the steady state that bifurcates is a pair of domain-filling rolls, each roll having $\Gamma_r=10$. States with two pairs of rolls ($\Gamma_r=5$) bifurcate at $R\approx730.7$, those with three pairs of rolls bifurcate at $R\approx744.5$, and so on. When $R$ is large enough for several such roll states to exist, but still small enough that $R-R_L\ll R_L$, the asymptotic analysis of \citet{Chapman1980} suggests that only the widest pair of rolls will be dynamically stable.

\subsection{Computational implementation}
\label{sec: implementation}

We carried out DNS of the equations~\cref{eq: bouss} using the code Dedalus v3 \citep{Burns_2020}, which has been extensively verified. In fact, the Dedalus codebase includes an example script for \mbox{2-D} RBC that we modified for our purposes. In all simulations, the spatial domain is discretized spectrally with 24 Chebyshev modes in the vertical direction and $24\Gamma$ Fourier modes in the horizontal. We find domain-filling rolls only at modest $R$ values, as reported in the next section, and at these $R$ our resolution is similar to or slightly lower than that in the spectral simulations of \citet{Johnston2009} (personal communication). For time stepping we use Dedalus's RK222 method, which implements the implicit--explicit Runge--Kutta scheme described in section 2.6 of \citet{Ascher_1997}. In simulations with fixed $R$ and $Pr$, the diffusion terms are treated implicitly in the time stepping. In some simulations we change $R$ and $Pr$ at each time step, for reasons explained later, and in these cases diffusion terms are treated explicitly because changing implicit terms requires recomputing matrix decompositions. Correctness of our implementation was verified by carrying out simulations with $(\Gamma,Pr)=(2,1)$ at certain small $R$ values where the flow converges to a steady state, and confirming that the mean boundary temperature difference $\delta\overline{T}$ agreed with values from \citet{Johnston2009} to within 0.02\% (personal communication).

\subsection{Parameter exploration}
\label{sec: exploration}

Carrying out DNS of a persistent pair of domain-filling rolls is possible only at $(R,Pr)$ values for which such a state is locally attracting, and it requires initial conditions that lie inside this attractor's basin of attraction. One might guess at such initial conditions by choosing a simple velocity field that resembles rolls with the right periodicity, as done in the study of \citet{Wang_2020} with fixed-temperature boundaries, but we find this strategy insufficient for our present aim. \citet{Wang_2020} used sinusoidal initial conditions with $n$ pairs of rolls (for various $n$) and found multiple locally stable states that each resulted from a different $n$ value. Here, however, sinusoidal initial conditions with $n=1$ often do not lead to domain-filling rolls at the end of the DNS, including in cases where we find such rolls by other means. Thus, even at $(R,Pr)$ where domain-filling rolls are an attracting state, the basin of attraction may be small in some sense, and care is needed to construct an initial condition in this basin.

At each new $(R,Pr)$ pair, our strategy to find an attracting state with domain-filling rolls is to start with such a state already found at a nearby $(R,Pr)$ pair. Finding a first state of this type can be done easily by choosing $R$ just slightly above the linear stability threshold $R_L$ because, in this weakly nonlinear regime, steady domain-filling rolls are the unique stable state \citep{Chapman1980}. After domain-filling rolls have developed in DNS, one can continue the DNS while changing $R$ and/or $Pr$ continuously in time---i.e., at each time step. In our DNS we increase or decrease $R$ at the rate $\tfrac{{\rm d}}{{\rm d}t}R=\pm 0.02\,R$, and likewise for $Pr$, and then hold parameters constant once the desired values are reached. The constant 0.02 was chosen based on trial and error; there were cases where we maintained a two-roll state using 0.02 but not with a larger constant, whereas we did not find cases where 0.02 failed but a smaller constant succeeded. We have also maintained two-roll states while changing $\Gamma$---e.g., we started with $\Gamma=2$, where two-roll states are ubiquitous \citep{Johnston2009}, and gradually increased $\Gamma$ to 20. This also succeeded, but to explore the $R$--$Pr$ plane it suffices to fix $\Gamma=20$.

Each time $(R,Pr)$ reaches desired values in DNS while maintaining a two-roll state, we fix the parameters and continue the DNS, switching to implicit treatment of the diffusion terms in the time stepping so that larger steps are numerically stable. If the DNS eventually approaches a steady state (as detected via vanishing time derivatives) with one pair of rolls, we conclude local stability of this steady state and halt the simulation. If the simulation remains time-dependent, as in the example rolls of \cref{fig: rolls}, we continue the DNS. As time goes on, this time-dependent two-roll state will either persist or break up into four rolls. To detect such break-up, we found it simplest to monitor certain instantaneous quantities by eye at regular time intervals, rather than fully automating the process. The three quantities we used were the instantaneous temperature field (as in \cref{fig: rolls}), the maximum of this field over the vertical direction (plotted versus $x$) and streamlines of the flow. With these three quantities, we manually classified whether the flow remained in a two-roll state. Simulations were stopped after two-roll states either (1) broke up into four rolls or (2) persisted at fixed $(R,Pr)$ for at least $10\Gamma/u_{rms}$ freefall time units, where $u_{rms}$ is the root-mean-square value of the dimensionless velocity field $\bu$. The criterion of $10\Gamma/u_{rms}$ freefall times is chosen as roughly the time for fluid particles to circumnavigate a roll 10 times: this distance is approximately $\Gamma$, and the velocity is roughly $u_{rms}$ away from the boundaries. Some simulations were run for longer, and in no cases did we find two-roll states persisting for a time of $10\Gamma/u_{rms}$ and then breaking up later.

\section{Results}
\label{sec: res}

\begin{figure}
\centering 
\begin{tikzpicture}
\node[anchor=south west, inner sep=0] (img) at (0,0)
{\includegraphics[trim={28 30 10 10},clip,width=0.6\textwidth]{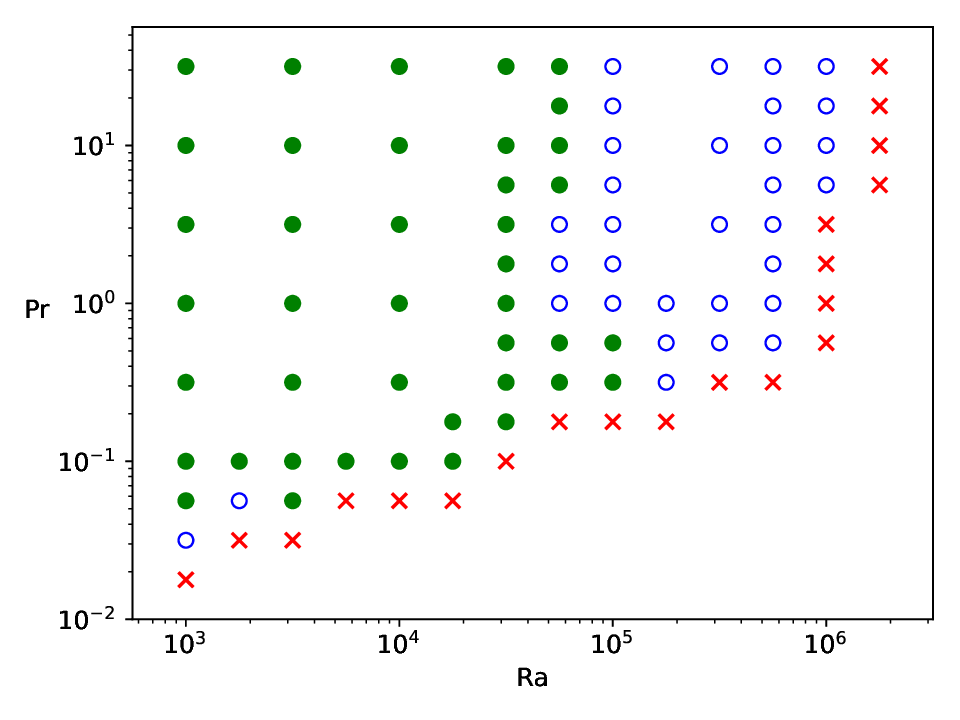}};
\begin{scope}[x={(img.south east)}, y={(img.north west)}]
\node at (0.5,-0.05) {$R$};
\node at (-0.03,0.5) {$Pr$};
\end{scope}
\end{tikzpicture}
\caption{Pairs of $(R,Pr)$ at which the DNS procedure described in the text did (circles) and did not~({\color{red}$\times$}) find a single pair of convection rolls that fill the domain of aspect ratio $\Gamma=20$. In DNS where the rolls persist, they either converge towards a steady flow ({\color{mygreen}$\bullet$}) or remain unsteady~({\color{blue}$\circ$}).}
\label{fig: plane}
\end{figure}

\Cref{fig: plane} shows where in the $R$--$Pr$ plane we found states with a single pair of domain-filling rolls. 
Each circle denotes parameter values at which our DNS lead to a roll pair that is either steady and stable~({\color{mygreen}$\bullet$}) or unsteady but persistent~({\color{blue}$\circ$}). Values at which we could not find two-roll states are also indicated~({\color{red}$\times$}). The parameter values are logarithmically spaced with a resolution of $10^{1/4}$ or $10^{1/2}$, with the finer spacing used near boundaries of parameter regimes. Aside from some of the lowest-$R$ cases, where two-roll states are easily found, each simulation represented in \cref{fig: plane} was initialized from an already-developed flow at adjacent parameter values. As described in \cref{sec: exploration}, parameters were changed continuously to their new values and then held there. This continuous change was able to maintain two-roll states in a number of cases where instantaneously changing $R$ or $Pr$ by a factor of $10^{1/4}$ (or even $10^{1/8}$) triggered a transition to a four-roll state. The points in \cref{fig: plane} marked by red crosses are where we could not maintain two-roll states when raising $R$ or decreasing $Pr$, including some cases where we tried changing the parameter values more slowly in time.

\Cref{fig: plane} suggests the approximate extent of the regime in the $R$--$Pr$ plane wherein domain-filling rolls are locally attracting. The red crosses constitute the approximate boundary between this regime and its complement. Farther into this complementary regime, at larger $R$ and smaller $Pr$, we made some additional attempts to find domain-filling rolls, but always without success. We have not carried out DNS at $Pr$ larger than $10^{3/2}$, but we speculate that the regime with domain-filling rolls extends to arbitrarily large $Pr$. The section of the regime boundary where rolls are steady could be found much more precisely by computing steady rolls using iterative methods, and then determining their linear stability. However, the part of the regime boundary where rolls are time-dependent would be much harder to find precisely. In general it is hard to determine parameters at which a chaotic attractor turns into a chaotic saddle with very long transients, even in low-dimensional ordinary differential equations. Furthermore, near the part of the regime boundary where the two-roll state is unsteady, it can even be ambiguous whether or not the DNS remains in this state because small recirculation zones can come in and out of existence. Nevertheless, the approximate shape of the regimes in \cref{fig: plane} already distinguishes our case from the others in \cref{tab: intro}.

\section{Conclusions}
\label{sec: con}

This study adds to, but is far from completing, the catalogue of scales on which superstructures can self-organize in different variants of RBC. The partial picture from previous studies, which is summarized by \cref{tab: intro} in the Introduction, shows the importance of dimension and boundary conditions but not of the dimensionless parameters. So long as the parameter values induce convection that is at least moderately strong, these previous studies found only weak influence of the Rayleigh and Prandtl numbers on scale selection. To complete the missing entry in the table, one must determine the range of scales that \mbox{2-D} rolls can have with fixed-flux boundaries. We have studied only larger scales, in particular by trying to find domain-filling rolls. Our main result is that such rolls can be found only when $R$ is not too large and $Pr$ is not too small. The possible scales of superstructures therefore depend on the parameters much more strongly than in previously studied cases. Regardless, as the picture of possible scales clarifies, the larger challenge is to understand the mechanisms that promote or prohibit different scales in different cases. 

With fixed-temperature boundaries, there are several partial explanations for the observed scales. In no-slip cases, impossibility of very wide superstructures can be understood via horizontal temperature gradients. Since fluid temperatures must stay between the boundary values, the largest horizontal gradients that can be sustained across a superstructure become weaker as the structures widen. When horizontal gradients are too weak, they cannot sustain horizontal flow against the drag of no-slip boundaries. In stress-free cases, on the other hand, domain-filling structures are possible in \mbox{2-D}, and their absence in \mbox{3-D} requires another explanation. In \mbox{3-D}, with either velocity boundary condition, it may be more fruitful to study why the flow selects particular $O(1)$ scales. Past authors have approached this question by examining perturbations to the turbulent state, including leading Lyapunov vectors \citep{Vieweg_2021a} and modes with maximum linear amplification around the turbulent mean \citep{Zhou_2026}. Such methods of analysis provide insight \emph{a posteriori}, but they do not give a way to predict scale selection \emph{a priori}, so theoretical questions remain.

With fixed-flux boundaries, fewer explanations have been proposed for domain-filling superstructures. For \mbox{2-D} rolls when $R$ is only slightly above the value where the conductive state becomes unstable, asymptotics show that non-domain-filling rolls are unstable to longer wavelengths \citep{Chapman1980}, which prompts the rolls to coalesce until there is one domain-filling pair. It also has been argued, based on leading Lyapunov vectors, that similar instability to longer wavelengths causes superstructures to coalesce in turbulent \mbox{3-D} convection \citep{Vieweg_2021a}. However, our new findings offer a way to test the validity of this explanation---or any other proposed explanation---for why fixed-flux boundaries create domain-filling superstructures in \mbox{3-D}. In particular, any such explanation should be able to predict when domain-filling rolls do or do not persist in \mbox{2-D} at various parameter values. Furthermore, our findings are also a way to test the explanation of \citet{Wang_2020} that wide \mbox{2-D} rolls are not found---in the fixed-temperature, no-slip case---due to elliptical instability. If this view is predictive, it should explain why we find domain-filling rolls only at certain parameter values.

In all cases where domain-filling superstructures have been found in large domains, even larger domains remain to be explored. In the \mbox{2-D}, arbitrarily wide rolls are expected in the weakly supercritical regime with fixed-flux boundaries, regardless of the velocity boundary conditions. Beyond this regime, we have found rolls filling a domain of aspect ratio 20 in the no-slip case, but it is not known whether these persist in even wider domains, or with stress-free boundary conditions, although the latter seems likely. With stress-free but fixed-temperature boundaries, domain-filling roll pairs have been found in domains of aspect ratio up to 128. In the \mbox{2-D} stress-free cases, it is an open question whether a single pair of rolls can be arbitrarily wide, and whether this limit of rolls quantitatively approaches the `windy convection' state \citep{Goluskin2014,Wang_2020b} that indeed resembles wide rolls away from their plumes. Similarly, the question of whether arbitrarily wide rolls persist may be related to the same question about windy convection in wide domains, whose answer is murky due to extremely long-lived transients \citep{Wang_2023}. In the \mbox{3-D} fixed-flux case, the superstructures that have been found are quite large, filling domains whose period in each horizontal direction is 60 times the layer height. We speculate that such superstructures can be arbitrarily large with stress-free boundaries, but that their scale may saturate on even wider domains with no-slip boundaries. Such speculations call for further study in domains of extreme horizontal extent, pushing the limits of what is possible by either DNS or laboratory experiment.

\section*{Acknowledgements}
We thank Philipp Vieweg for helpful discussion. All authors were supported by the NSERC Discovery Grants Program via award numbers RGPIN-2018-04263, RGPAS-2018-522657, DGECR-2018-0037 and RGPIN-2025-06823. The first author was supported by an NSERC Undergraduate Student Research Award, and the second author was supported by a Mitacs Research Training Award.

\section*{Declaration of interests}

The authors report no conflict of interest.

\newpage
\bibliographystyle{jfm}
\bibliography{refs.bib}

\end{document}